\documentclass{ws-procs9x6}
\usepackage{epsfig}

\usepackage{graphicx}
\usepackage[figuresright]{rotating}

\newcommand{\AmS}{{\protect\the\textfont2
  A\kern-.1667em\lower.5ex\hbox{M}\kern-.125emS}}

\textfloatsep=\floatsep           
\intextsep=\floatsep

\title{Present status of IGEX dark matter search at Canfranc Underground Laboratory}

\author{
I.G.~Irastorza$^{\lowercase{a}}$\footnote{\lowercase{Attending
speaker, E-mail:
Igor.Irastorza@cern.ch}}~\footnote{\uppercase{P}resent address:
\uppercase{CERN, EP D}ivision, \uppercase{CH}-1211
\uppercase{G}eneva 23, \uppercase{S}witzerland},
A.~Morales$^{\lowercase{a}}$, C.E.~Aalseth$^{\lowercase{b}}$,
F.T.~Avignone~III$^{\lowercase{b}}$,
R.L.~Brodzinski$^{\lowercase{c}}$,
J.~M.~Carmona$^{\lowercase{a}}$, S.~Cebri\'{a}n$^{\lowercase{a}}$,
E.~Garc\'{\i}a$^{\lowercase{a}}$,
I.V.~Kirpichnikov$^{\lowercase{d}}$,A.A.~Klimenko$^{\lowercase{e}}$,
G.~Luz\'{o}n$^{\lowercase{a}}$, H.S.~Miley$^{\lowercase{c}}$,
J.~Morales$^{\lowercase{a}}$,
A.~Ortiz~de~Sol\'{o}rzano$^{\lowercase{a}}$,
S.B.~Osetrov$^{\lowercase{e}}$, V.S.~Pogosov$^{\lowercase{f}}$,
J.~Puimed\'{o}n$^{\lowercase{a}}$, J.H.~Reeves$^{\lowercase{c}}$,
M.L.~Sarsa$^{\lowercase{a}}$,A.A.~Smolnikov$^{\lowercase{e}}$,
A.G.~Tamanyan$^{\lowercase{f}}$,
 A.A.~Vasenko$^{\lowercase{e}}$,
S.I.~Vasiliev$^{\lowercase{e}}$, J.A.~Villar$^{\lowercase{a}}$}

\address{The IGEX Collaboration\\ ~\\
$^{a}$Laboratory of Nuclear and High Energy Physics, University of
Zaragoza, 50009 Zaragoza, Spain
\\
$^{b}$University of South Carolina, Columbia, South Carolina 29208
USA
\\
$^{c}$Pacific Northwest National Laboratory, Richland, Washington
99352 USA
\\
$^{d}$Institute for Theoretical and Experimental Physics, 117 259
Moscow, Russia
\\
$^{e}$Institute for Nuclear Research, Baksan Neutrino Observatory,
361 609 Neutrino, Russia
\\
$^{f}$Yerevan Physical Institute, 375 036 Yerevan, Armenia }

\begin{document}

\maketitle

\abstracts{One IGEX $^{76}$Ge double-beta decay detector is
currently operating in the Canfranc Underground Laboratory in a
search for dark matter WIMPs, through the Ge nuclear recoil
produced by the WIMP elastic scattering. In this talk we report on
the on-going efforts to understand and eventually reject the
background at low energy. These efforts have led to the
improvement of the neutron shielding and to partial reduction of
the background, but still the remaining events are not totally
identified. A tritium contamination or muon-induced neutrons are
considered as possible sources, simulations and experimental test
being still under progress. According to the success of this study
we comment the prospects of the experiment as well as those of its
future extension, the GEDEON dark matter experiment.}


\section{Introduction}

Recent cosmological observations and robust theoretical arguments
require an important Dark Matter component ($\Omega_{DM}\sim
25-30\%$) in our universe, which is supposed to be made mostly of
non-baryonic particles. Weakly Interacting Massive (and neutral)
Particles (WIMPs), which are favourite candidates to such
non-baryonic component, could fill the galactic halos accounting
for the flat rotation curves which are measured for many galaxies.
They could be detected by measuring the nuclear recoil produced by
their elastic scattering off target nuclei in a suitable
detector\cite{MoralesTAUP2001}. The IGEX dark matter experiment is
currently operating one germanium detector with this purpose in
the Canfranc Underground Laboratory. In this talk we review the
present status of the experiment. Following a brief description in
section 2, we focus on the on-going work about the identification
and rejection of the low energy background in the section 3. In
section 4 we will present the prospects of the experiment, as well
as of the GEDEON project, a proposed extension of the IGEX dark
matter search that will be commented in some detail. We will
finish with the conclusions in section 5.

\section{Experiment}
The IGEX experiment \cite{Aal,Gon99}, optimized for detecting
$^{76}$Ge double-beta decay, has been described in detail
elsewhere. One of the IGEX detectors of 2.2 kg (active mass $\sim$
2.0 kg), enriched up to 86 \% in $^{76}$Ge, is being used to look
for WIMPs interacting coherently with the germanium nuclei. The Ge
detector and its cryostat were fabricated following
state-of-the-art ultralow background techniques and using only
selected radiopure material components (see Ref \cite{Aal,Mor00}).

The detector shielding has been modified several times since the
beginning of the dark matter phase of IGEX. The improvements
concern basically the neutron moderator, which thickness went from
the 20 cms of polyethylene for the first published data
\cite{Mor00} to 40 cms (of polyethylene and borated water tanks)
in \cite{Morales:2001} and to 80 cms in the more recent data
taken. After the first changes of the shielding, it much better
covers the whole set-up, due to the removal of the other detectors
(RG-I, RG-III and COSME), which dewars did not allow us to
perfectly close the polyethylene wall. These changes were
motivated by a previous study of the possible sources of
background in IGEX based on preliminary simulations, which pointed
out that the neutrons from the surrounding rock could contribute
considerably to the low energy background. This background study
is still going on as commented in the next section. For more
details on the IGEX shielding we refer to \cite{Mor00} and
\cite{Morales:2001}.

In addition to the data acquisition system used in the first runs
(described in \cite{Mor00}), a specific pulse shape analysis was
implemented for subsequents sets of data\cite{Morales:2001}. The
charge pulse shapes of each event before and after amplification
are recorded by two 800 MHz LeCroy 9362 digital scopes. These are
analyzed one by one by means of a method based on wavelet
techniques\cite{wavelets} which allows us to get the probability
of this pulse to have been produced by a random fluctuation of the
baseline. This probability is used as a criterium to reject events
coming from electronic noise or microphonics. According to the
calibration of the method, it works very efficiently for events
above 4 keV.

\begin{figure}[t]
\centerline{ \epsfxsize=9cm \epsffile{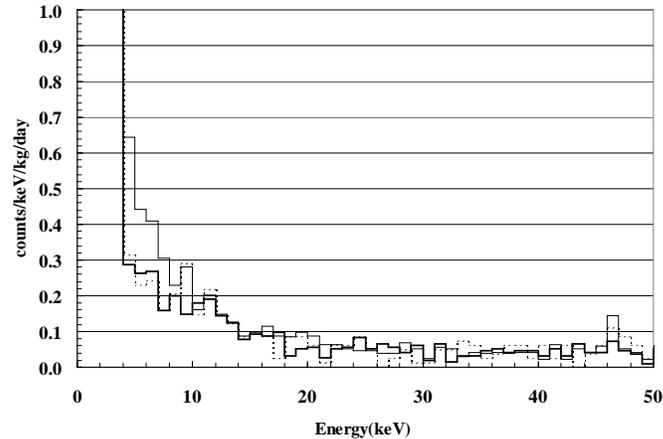} }
 \caption{\footnotesize Normalized low energy spectrum of the IGEX RG-II detector
 corresponding to the three different neutron shielding conditions mentioned in the
 text: 236~kg~d with 20 cms of polyethylene (thin solid line), 194~kg~d with 40 cms of
 polyethylene and borated water (thick solid line) and 82~kg~day with 80 cms of polyethylene and borated
 water (dashed line).}
 \label{dm-ig-1}
\end{figure}

\begin{figure}[t]
\centerline{ \epsfxsize=7.5cm \epsffile{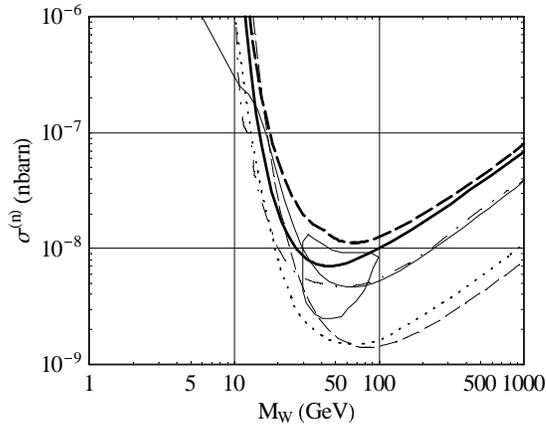} }
 \caption{\footnotesize IGEX-DM exclusion plot for spin-independent
interaction obtained with the data before (thick dashed line) and
after (thick solid line) the neutron shielding improvement,
compared with the results from other experiments which exclusions
also enter in the DAMA region:
CDMS\protect\cite{Abusaidi:2000}(thin dot-dashed line),
EDELWEISS\protect\cite{Benoit:2001}(thin dashed line). Also
included is the very recent curve presented by the ZEPLIN
collaboration in this same conference\protect\cite{zeplin} (dotted
line). The thin solid line is the exclusion line obtained by DAMA
NaI-0 by using Pulse Shape Discrimination\protect\cite{Ber96}. The
closed line corresponds to the (3$\sigma$) annual modulation
effect reported by the DAMA collaboration (including NaI-1,2,3,4
runnings)\protect\cite{Ber99}.
} \label{dm-ig-2}
\end{figure}

\section{Status}

As has been stressed in previous papers\cite{Mor00}, the
sensitivity of the experiment can be substantially improved if
modest reduction of background at low energies is achieved.
Currently the main concern of the collaboration is the
identification of the low energy background in order to design
strategies that allow us to reduce it. Several studies based on
simulation and experimentation are still under work, but some
improvements in that sense have been already achieved and recently
published\cite{Morales:2001}. In Fig.~\ref{dm-ig-1} we shown the
spectra taken before (thin line) and after (thick line) the first
improvement of the neutron shielding. Those spectra are the same
as the ones published in \cite{Mor00} and \cite{Morales:2001}
respectively, but with more statistics added. The simulations that
motivated this improvement are in qualitative agreement with the
reduction of the experimental background actually obtained (that
amounts to a 50\% in the first energy bins) although the remaining
low energy population of events is not yet fully identified. More
recently data has been taken with an even thicker neutron
moderator wall (80 cms), obtaining a similar level of background,
as plotted in Fig.~\ref{dm-ig-1} (dotted line). The fact that this
second change does not improve the background is also consistent
with our simulations and with the fact the neutrons from the rock
have been practically shielded out from the IGEX background. On
the other hand, simulations taking into account surrounding
contaminations of typical radio-isotopes ($^{60}$Co, $^{40}$K,
$^{238}$U/$^{222}$Rn, $^{232}$Th, $^{210}$Pb,...) only explain a
flat background of $\sim$0.04~c/keV/kg/d as the one experimentally
obtained above $\sim$20 keV. The identification of the remaining
events that populate the low energy region above this level
(accounting for up to $\sim$0.2 c/kev/kg/d between 4 and 10 keV)
is currently the main goal of the experiment. One possibility
under consideration for the origin of these events is an internal
contamination of tritium. On the other hand, the effect of
muon-induced neutrons is being studied in detail, the preliminary
conclusion being that they can hardly explain this population. Let
us say that all these studies will allow also to set the prospects
of the projected extension of IGEX for dark matter searches, the
GEDEON project.



The exclusion plots are derived from the both recorded spectra
(before and after the first modification) following the same set
of hypothesis and parameters used in previous papers (see
\cite{Mor00}) and are shown in Fig.~\ref{dm-ig-2} (thick solid and
dashed lines). They are compared with other recent exclusions.

\section{Prospects}

\begin{figure}[t]
\centerline{ \epsfxsize=7.5cm \epsffile{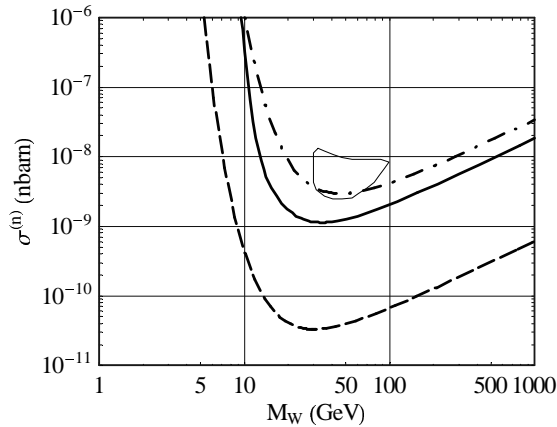} }
 \caption{\footnotesize IGEX-DM projections are shown for
 a flat background rate of 0.1~c/keV/kg/day (dot-dashed line) and 0.04~c/keV/kg/day (solid line) down
 to the threshold at 4 keV, for 1~kg~year of exposure.
 The exclusion contour expected for GEDEON is also
 shown (dashed line) as explained in the text.} \label{prospects}
\end{figure}

As stated before, background identification works are still in
progress and some strategies are being considered to further
reduce the low energy background. If this reduction is achieved,
very interesting perspectives can be set for IGEX. In
Fig.~\ref{prospects} we plot the exclusions obtained with a flat
background of 0.1 c/kg/keV/day (dot-dashed line) and of 0.04
c/kg/keV/day (solid line) down to the current 4 keV threshold for
an exposure of 1~kg~year. In particular, the complete DAMA region
could be tested with a moderate improvement of the IGEX
performances. The dashed line in Fig.~\ref{prospects} corresponds
to a flat background of 0.002 c/kg/keV/day down to a threshold of
4 keV and 24 kg y of exposure, which are the parameters expected
for GEDEON (GErmanium DEtectors in ONe cryostat), a new
experimental project on WIMP detection using larger masses of
natural germanium planned as an extension of the IGEX dark matter
search (see ref.~\cite{Morales:2001}). GEDEON would be massive
enough \cite{Cebrian:2001} to search also for the WIMP annual
modulation effect and explore positively an important part of the
WIMP parameter space including the DAMA region.

\section{Conclusions}

The IGEX dark matter experiment is carrying a thorough study of
its low energy experimental background which after the last
improvements of the neutron shielding is already at the level of
$\sim$0.2 c/keV/kg/day (between 4 and 10 keV), the lowest low
energy raw background achieved up to now. The possibility of
further background rejection sets interesting prospects for the
experiment. The GEDEON project, conceived as an natural extension
of the IGEX-DM experiment will have a substantially improved
background and will be able to explore the annual modulation
signature.

%

\end{document}